\documentclass[article,shortnames]{jss}

\usepackage{epsfig}
\usepackage{amsfonts}
\usepackage{amsmath}
\usepackage{gensymb}

\author{Ioanna Manolopoulou\\University College London \And 
Axel Hille\\Institute of Applied Statistics\\ Dr J\"org Schnitker\And
Brent Emerson\\ Instituto de \\Productos Naturales\\ y Agrobiologia, and\\University of East Anglia}
\title{\pkg{BPEC}: An \proglang{R} Package for Bayesian Phylogeographic and Ecological Clustering}

\Plainauthor{Ioanna Manolopoulou, Axel Hille} %% comma-separated
\Plaintitle{BPEC: An R Package for Bayesian Phylogeographic and Ecological Clustering} %% without formatting
\Shorttitle{Phylogeographic Ecological Clustering} %% a short title (if necessary)

\Abstract{
\pkg{BPEC} is an \proglang{R} package for Bayesian Phylogeographic and Ecological Clustering which allows geographical, environmental and phenotypic measurements to be combined with DNA sequences in order to reveal geographic structuring of DNA sequence clusters consistent with migration events. DNA sequences are modelled using a collapsed version of a simplified coalescent model projected onto haplotype trees, which subsequently give rise to constrained clusterings as migrations occur. Within each cluster, a multivariate Gaussian distribution of the covariates (geographical, environmental, phenotypic) is used. 
Inference follows tailored Reversible Jump Markov chain Monte Carlo sampling so that the number of clusters (i.e., migrations) does not need to be pre-specified. A number of output plots and visualizations are provided which reflect the posterior distribution of the parameters of interest. \pkg{BPEC} also includes functions that create output files which can be loaded into Google Earth. The package commands are illustrated through an example dataset of the polytypic Near Eastern brown frog \emph{Rana macrocnemis} analysed using \pkg{BPEC}.
}
\Plainkeywords{biogeography, population genetics, Bayesian computation, R} %% without formatting
\Address{
Ioanna Manolopoulou\\
Department of Statistical Science\
University College London\\
London WC1E 6BT, UK\\
E-mail: \email{ioanna@stats.ucl.ac.uk}\\
URL: \url{http://www.ucl.ac.uk/~ucakima/}\\

Axel Hille\\
Institute of Applied Statistics Dr.~J\"org Schnitker Ltd. \\
Oberntorwall 16-18 \\
D-33602 Bielefeld, Germany\\
E-mail: \email{axel.hille@gmx.net}\\

Brent C. Emerson\\
Island Ecology and Evolution Research Group\\ 
Instituto de Productos Naturales y Agrobiologia\\ 
C/Astrof\i sico Francisco Sanchez 3\\
La Laguna\\
Tenerife\\ 
Canary Islands 38206\\
Spain, and\\  
School of Biological Sciences\\
University of East Anglia\\
Norwich Research Park\\
Norwich NR4 7TJ\\
UK\\
E-mail: \email{bemerson@ipna.csic.es}
}

\begin{document}

\maketitle

\section{Introduction}
Phylogeography can be considered the nexus between classical population genetics, phylogenetics and historical biogeography, with much conceptual and analytical overlap with all three, but particularly with  population genetics. Phylogeography was born from the integration of population genetics and phylogenetics to work at the micro- macroevolutionary interface \citep{hickerson2010phylogeography}, being an evolved discipline that seeks to integrate the genealogical relationships among DNA lineages (sequences) with their geographic distributions to infer historical events that have shaped the contemporary distributions of species and their genetic variation. However, while population genetics, phylogenetics and historical biogeography have witnessed a growth of analytical approaches in recent years, there has been a relative dearth of analytical approaches within the field of phylogeography, with several reviews summarising these \citep[e.g.][]{knowles2009statistical,bloomquist2010three,hickerson2010phylogeography}. To place our work into a broader context, we provide a brief summary of the state of the art within the field of phylogeography, but the aforementioned reviews should be referred to for more detail.

Historical biogeography seeks to understand the processes that have shaped the evolution of geographic differences among related species (i.e.,~interspecific process), and may involve timescales that extend back tens of million of years or more. In contrast, phylogeography concerns both the quantification of the geographic structuring of genetic variation within species, and understanding the process that has shaped said structure (i.e.~intraspecific process). Thus phylogeographic analyses typically involve time-scales that don't extend back more than a few million years. A challenge for phylogeographic analysis is to simultaneously account for evolutionary processes over spatial and temporal dimensions, and perhaps for this reason the phylogeographer's toolkit is a mixed bag of approaches encompassing various objectives within this framework. Some population genetic methods find relevance in phylogeography, precisely because they do not use geographical information explicitly, but rely on population genetics modelling to infer the geography of structure. For example, \pkg{STRUCTURE} \citep{pritchard2000inference} infers population structure purely from genotype data through a Latent Dirichlet Allocation model. Population subdivisions are assessed on the basis of multi-locus allele frequencies which are directly learnt from data. More recently, \cite{jombart2010discriminant}  developed \pkg{DAPC}, a principal-components alternative to \pkg{STRUCTURE} which can computationally efficiently deal with large amounts of data. In these approaches one describes genetic groupings in the absence of spatial information, onto which phylogeographic inferences can then be conditioned. Fully model-based extensions of spatially-explicit inferences of population structure such as \pkg{GENELAND} \citep{guillot2005geneland} and \cite{cheng2013hierarchical} assume that the spatial domain occupied by the inferred clusters can be approximated by a small number of polygons based on Voronoi tessellations. Drawing inferences about these cluster domains (and thus about cluster membership) amounts to inferring the location and cluster memberships of the polygons. Finally, recent approaches such as \cite{jay2015pops} introduced spatially-dependent cluster membership probabilities through a regression model. These approaches use multilocus genotype data for the inference of spatial genetic structure, and therefore the absence of a coalescent framework limits inferences across the temporal dimension. 

Methods that use the evolutionary relationships among alleles for phylogeographic analysis open the door for jointly investigating the spatial and temporal dimensions of genetic relatedness among individuals. Early phylogeography relied upon qualitative assessments of the geographic relationships within a gene genealogy, together with estimated dates of gene tree branching events. In this approach demography was directly inferred from the phylogenetic relationships of alleles, with limited importance given to the potentially confounding effects of coalescent stochasticity \citep{hickerson2010phylogeography}. Such stochasticity could give rise to similarly probable alternative demographic explanations for a given data set. To address this, simulation-based statistical methods based on coalescent models for parameter estimation have emerged giving rise to statistical phylogeography \citep{knowles2002statistical,knowles2009statistical} allowing for testing among competing demographic models.

With regard to the joint analysis of the genealogical and spatial relationships of DNA sequences, we are only aware of two implementations to date.  \cite[][]{lemey2009bayesian} developed a  fully model-based Bayesian phylogeographic inference framework, assuming a diffusion model for the geographical migration of nodes on a phylogenetic tree, so that evolution and migration events occur in a continuous-time framework. More recently, \cite{guindon2016demographic} modelled spatial distribution as a gradual dispersal across a continuous landscape. 

Here we present a novel \proglang{R}  \citep{RCRANcitation2017} package available from the Comprehensive \proglang{R} Archive Network at \url{http://CRAN.R-project.org/package=BPEC} which automates Bayesian Phylogeographic and Ecological Clustering (\pkg{BPEC}) analysis \citep{manolopoulou2011bayesian,manolopoulou2012phylogeographic}. \pkg{BPEC} is a model-based approach which assumes that population substructure is the result of individuals migrating into a new area  (i.e.~dispersal). \pkg{BPEC} differs from the methods of \cite{lemey2009bayesian} in that it explicitly models geographical ranges, assuming that sampling localities are random samples from the entire landscape. In contrast to the continuous approach of \cite{guindon2016demographic}, it addresses the phylogeographic structure by inferring geographically structured clusters of DNA sequences as the result of distinct colonisation events, while also admitting a model for the evolutionary history. Here a cluster is defined as a subnetwork of sequences within the haplotype tree that are geographically aggregated and have similar ecological characteristics. \pkg{BPEC} performs full Bayesian inference, which means that it provides an entire posterior distribution over phylogeographic clusterings; although this comes at a computational cost, the ability to provide uncertainty measures is valuable in terms of understanding the impact on scientific hypotheses of interest.

The key function of \pkg{BPEC} inputs non-recombinant DNA sequences and geographical locations, as well as any additional covariates available, such as temperature or phenotypic characteristics, in order to identify clusters that are consistent with migration. The results of the analysis provide estimates on the number of migration events, the geographical distribution of the clusters, ancestral locations and clustered tree structure. Aside from providing estimates for the quantities of interest, \pkg{BPEC} also provides measures of uncertainty of the conclusions and functions for post-processing. Finally, \pkg{BPEC} is supplemented with various visualization tools interfacing with geographical mapping resources to aid interpretation. In Section~\ref{sec:model}, we present the \pkg{BPEC} model, followed by the corresponding Bayesian computation methods in Section~\ref{sec:bayes}. Section~\ref{sec:brownfrog} describes an example dataset of the eastern lineages of polytypic Near Eastern brown frogs, \emph{Rana macrocnemis} (Boulenger, 1885), from the Caucasus region \citep{tarkhnishvili2001humid}, and the \proglang{R} user interface is presented in Section~\ref{sec:interface} through the analysis of the example dataset. The output is interpreted in Section~\ref{sec:analysis} and the paper concludes with a short discussion in Section~\ref{sec:discussion}.

\section{Model}\label{sec:model}
The aim of \pkg{BPEC} is to combine sequence data $\mathcal{S}$ with geographical and (optionally) ecological data $\mathcal{Y}$ for demographic inference regarding the geographic and (optionally) ecological structuring of genetic variation and thus potential geographical or ecological limitations to gene flow. To achieve this aim \pkg{BPEC} combines an evolutionary model for the genealogical relationships among sampled DNA sequences together with a geographical model representing dispersal events  forming clusters into a fully model-based framework.

\subsection{Haplotype tree model}
Approaches to model and estimate the evolutionary relationships among DNA sequences range from simple and elegant, such as the vanilla coalescent \citep{kingman} to complex with intractable~likelihood forms \citep{cornuet2014diyabc}. Questions such as the validity of a constant (or effectively constant) population size, independent nucleotide mutations, constant mutation across sites, time-dependence, presence of natural selection pressure, all play a role in defining an appropriate evolutionary model and have led to a variety of extensions of the basic model \citep{wakeley2013coalescent,hein2004gene}. In our case, typical datasets are expected to vary from a several hundred to no more than several thousand nucleotides, with low levels of polymorphism that typically characterise intraspecific data sets.

\begin{figure}[h!]
\centering
\includegraphics[width=10cm]{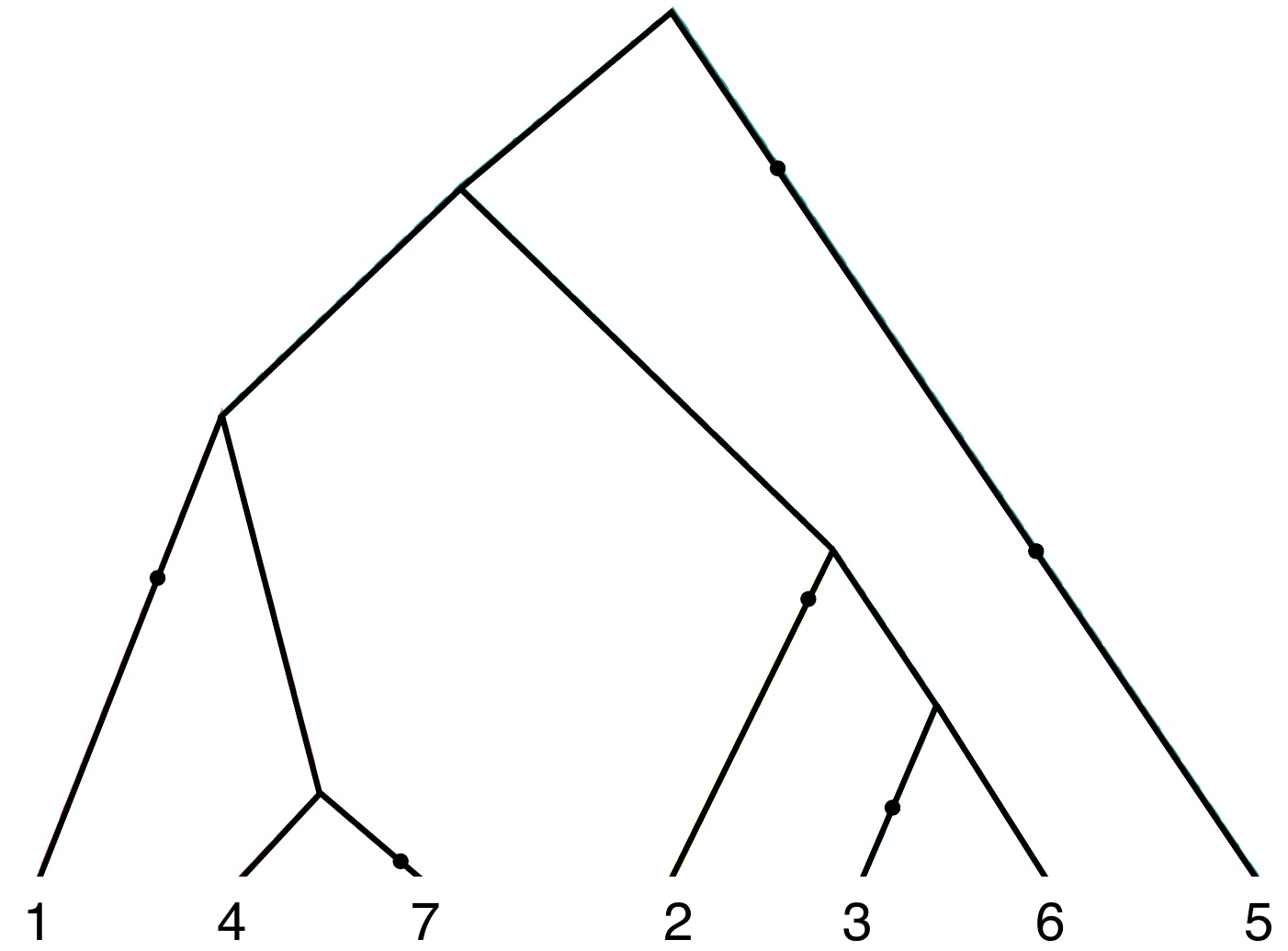}
\caption{ \label{coalescent}An example of a coalescent tree with mutations for 7 observed sequences including mutations; black dots represent mutations. Time evolves from top to bottom. }
\end{figure}

\begin{figure}[h!]
\centering
\includegraphics[width=5cm]{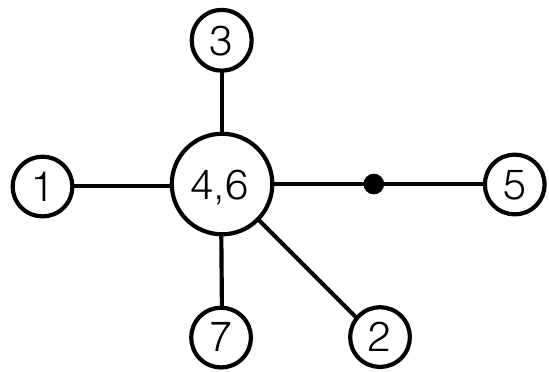}
\caption{\label{haplotree}The corresponding haplotype tree of Figure~\ref{coalescent}, where edges represent single effective mutations. The black dot represents an unobserved intermediate sequence. Note that sequences 4 and 6 correspond to the same haplotype.}
\end{figure}

As a result, the nucleotide data are noisy and often too weakly informative to allow for very complex models. The evolutionary relationships among a sample of DNA sequences can be represented in one of two ways: a coalescent tree or a haplotype tree or network. A coalescent tree is plotted against time and thus explicitly characterizes the most recent common ancestor. An example of a coalescent tree with mutations mapped on is shown in Figure~\ref{coalescent} where tips represent observed sequences and black circles indicate mutations, and the timing of the most recent common ancestors (MRCAs) among sequences is represented by branch lengths. In contrast, haplotype trees (Figure~\ref{haplotree}) summarise mutation differences among sampled sequences, so only implicitly carry information about time. To infer the root haplotype within a haplotype tree an evolutionary model is needed, but such models are not readily available. However, models such as the coalescent with mutations are available \citep{ethiergriffiths}.

There are two main tree representations of evolutionary histories. The first is via a coalescent tree (with or without mutations); the second is via cladograms, haplotype trees or networks. A coalescent tree is plotted against time and thus explicitly characterizes the most recent common ancestor. An example of a coalescent tree with mutations is shown in Figure~\ref{coalescent}: the tips represent observed sequences and black circles indicate mutations. As time progresses, both mutation and coalescence events occur.  In contrast, haplotype trees are plotted against number of mutations, so only implicitly carry information about time. The haplotype tree corresponding to the coalescent tree of Figure~\ref{coalescent} is shown in Figure~\ref{haplotree}. 
Given a haplotype tree, an evolutionary model is subsequently needed to allow us to draw inferences about the root haplotype within a given tree. However, probabilities of evolutionary histories over rooted haplotype trees are not readily available; models such as the coalescent with mutations are, instead, available \citep{ethiergriffiths}. 

A subtle complication derives from potential tree unidentifiability due to  repeated observations of the same haplotype that are to be expected when either or both the mutation rate and number of sampled nucleotides is insufficient to ascribe unique variation to all sample haplotypes. As an example, observations 4 and 6 in Figure~\ref{coalescent} correspond to the same haplotype, meaning that the 2 observations could be switched without having any effect on the likelihood of the tree. Observations may, however, be distinct with respect to the geographical or ecological information associated to each one. Aside from identifiability issues, exploring the space of equivalent trees requires cycling through a complex combinatorial object which quickly becomes computationally cumbersome. Collapsing sequences into haplotypes allows us to get around this issue, reducing the space of possible trees as exemplified in Figure~\ref{haplotree} where sequences 4 and 6 from the coalescent tree (Figure~\ref{coalescent}) are now represented by the same node.

In order to draw inferences about the haplotype tree, approaches can be fully model-based \citep{felsen3,mrb,drummond2012bayesian}, parsimony-based through the underlying tree \citep{rzhetsky,gascuel1}, or purely phenetic such as neighbour-joining or median-joining \citep{atteson1,gascuelsteel}. \pkg{BPEC} combines parsimonious approaches within a model-based framework.
Although an infinite set of haplotype or coalescent trees could be consistent with the sequence data $\mathcal{S}$, \pkg{BPEC} uses relaxed parsimony to reduce it to a finite set of `plausible' trees $\Omega$ represented via a graph \citep{manolopoulou2012phylogeographic}. The relaxed parsimony is defined by a threshold $d_s$ representing parsimony relaxation. Briefly, haplotypes are connected by an edge if they are a single mutation apart. When two groups of haplotypes are disconnected (with minimum mutation distance $d_{min}$), then any connection path with length up to $d_{min}+d_s$ is considered. The exact details of how to obtain $\Omega$ from $\mathcal{S}$ for a given $d_s$ can be found in algorithm A of \cite{manolopoulou2012phylogeographic}. This algorithm constructs a set of `realistic' trees  by cumulatively adding intermediate sequences following a relaxed parsimony assumption defined by the user-specified parsimony relaxation parameter  $d_s$.  In general, larger values of $d_s$ (up to a maximum value) yield more inclusive (and hence realistic) sets $\Omega$, but the choice of $d_s$ is often limited by computational power. 
For a fixed $d_s$, this algorithm inputs the DNA sequences at hand, and outputs a sequence network, including loops. The true haplotype tree is then assumed to be one  of the minimum spanning trees of this graph with equal probability and can be obtained through the breaking of loops.  

A  haplotype tree encodes less information than a coalescent tree with mutations. Firstly, a haplotype tree only encodes time through number of mutations. Secondly, it does not automatically define an ordering of events, starting from a root down to tips.  Even a rooted (i.e.,~one where the ancestral haplotype is specified) haplotype tree imposes only a partial ordering to the set of past mutation and coalescence events. 
Calculating probabilities over rooted haplotype trees therefore requires summing over all possibilities and orderings of past events given a temporal model; an example of possible orderings is shown in Appendix~\ref{appendix:temporal}. We denote a temporal ordering of events as $\mathcal{O}$, where $\mathcal{O}^\mathcal{S}_{r,T}$ denotes the set of all temporal orderings consistent with data $\mathcal{S}$ given a root $r$ and tree $T$ and assume that any temporal ordering of events is equally likely a priori. Conditionally on observed data (which restricts the possible trees to the space $\Omega$) this prior corresponds to a discrete uniform distribution over $\Omega$ and 
provides the following posterior probabilities for the root $r$ and tree $T$: 
\begin{eqnarray}
\mathbb{P}(r,T\mid \mathcal{S})&=& \frac{\left|\mathcal{O}^\mathcal{S}_{r,T}\right|}{\sum_{r,T}\left|\mathcal{O}^\mathcal{S}_{r,T}\right|},\nonumber
\end{eqnarray}
where $\left|\cdot\right|$ denotes the size of the set. 
Similarly, 
\begin{eqnarray}
\mathbb{P}(r\mid T,\mathcal{S})&=& \frac{\left|\mathcal{O}^\mathcal{S}_{r,T}\right|}{\sum_{r}\left|\mathcal{O}^\mathcal{S}_{r,T}\right|},\nonumber\\
\mathbb{P}(T\mid r,\mathcal{S})&=& \frac{\left|\mathcal{O}^\mathcal{S}_{r,T}\right|}{\sum_{T}\left|\mathcal{O}^\mathcal{S}_{r,T}\right|}.\nonumber
\end{eqnarray}

\begin{figure}[ht]
\centering
\includegraphics[width=4in]{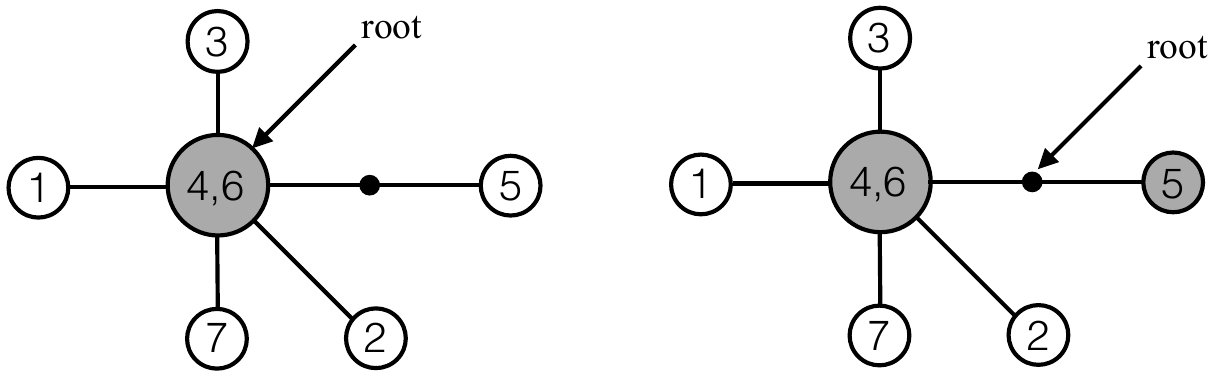}
\caption{\label{figancestral}Two possible root scenarios for the haplotype tree presented in Figure~\ref{haplotree}. In the left-hand panel, the root haplotype (shown in grey) is observed and thus any location will be inferred to be ancestral according to the proportion of observations of the inferred root haplotype 4. In the right-hand panel, the inferred root haplotype is not observed and the two equally divergent descendant haplotypes (shown in grey) are then used to infer ancestral locations as a function of the proportion of copies of either haplotype in a given location. }
\end{figure}

This model naturally takes into account the total number of combinations of mutational and coalescence events. Note that this model disregards the relative probability of coalescence versus mutation, essentially assuming that at every time point either are equally likely. The model can be extended to introduce a mutation rate $\theta$ (at the expense of computational complexity) which is simultaneously learnt and is used to refine the posterior probabilities of each tree. Although the haplotype tree model described provides a way of assigning posterior probabilities of haplotypes being ancestral, these need to be associated to sampling locations in order to infer the most ancestral location. \pkg{BPEC} assigns probabilities to each location based on the haplotypes observed in each. For each posterior sample, if the inferred root haplotype is observed, then each observed sequence that corresponds to that haplotype contributes equally to a location being ancestral. In other words, each location will be inferred to be ancestral with probability equal to the proportion of root haplotypes that were sampled in it. If the inferred root haplotype is not observed (i.e.~extinct or unsampled), then the oldest observed haplotypes derived from the inferred root are considered equally likely to be the `most ancestral' and thus each observation of one of these haplotypes contributes equally to the probability of each sampling location being ancestral. An example of this is shown in Figure~\ref{figancestral}. An important feature of this approach is that the probability of each location being ancestral depends on the proportional representation of each haplotype. This is to circumvent issues of wide sampling variability across locations.

\subsection{Clustering model}
\begin{figure}[h!]
\centering
\includegraphics[width=10cm]{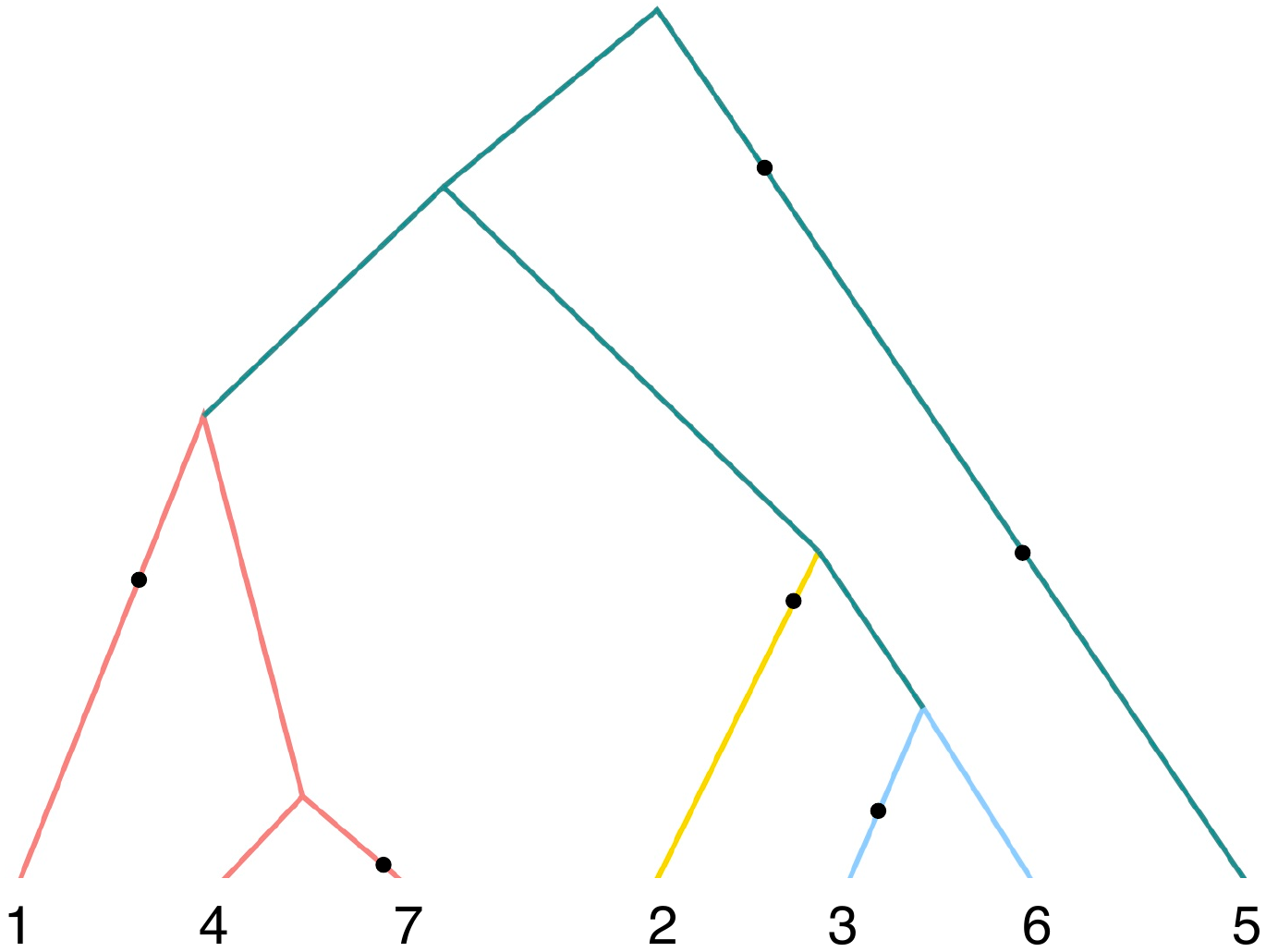}
\caption{\label{coalescentsubdivided}
Geographically informed genetic clustering of the coalescent tree from Figure~\ref{coalescent}. Different geographic clusters are represented by different colours. The inferred ancestral geographic area is represented by green with three migration events inferred to give rise to three derived geographic clusters.}
\end{figure}

The two main requirements to infer migration events for a given tree are: (i) a model for constructing constrained clusterings conditionally on a haplotype tree, and (ii) a model for the distribution of data within each cluster. A key assumption in our model is that new clusters are formed through the migration/dispersal/colonisation of a single individual (haplotype) founding a new geographically distinct cluster \citep{deiorio1,deiorio2}. All subsequent descendants of this founding haplotype belong to the new cluster, unless they migrate again. Given an inferred tree representing the genealogy, possible clusterings of the data are thus constrained by the tree while at the same time informed via the  geographic distribution (and optionally ecological data) of the observations for each individual. Figure~\ref{coalescentsubdivided} provides an illustration using the hypothetical coalescent tree of Figure~\ref{coalescent}.

\begin{figure}[h!]
\centering
\includegraphics[width=5cm]{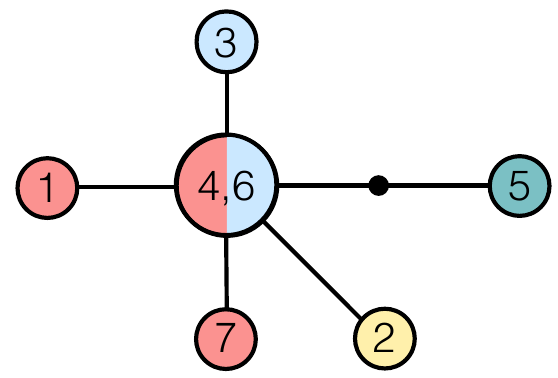}
\caption{\label{haplotreesubdivided} The clustered haplotype tree corresponding to the subdivided coalescent tree of Figure~\ref{coalescentsubdivided} where colour corresponds to cluster and size of node to the number of individuals sampled with each sequence. Edges represent single effective mutations and black dots represent unobserved intermediate haplotypes.
}
\end{figure}

The coalescent tree determines a set of constrained clusterings which are feasible through migration events. For example, observation 2, 3 and 6 in Figure~\ref{coalescentsubdivided} could have formed a single cluster together, but 6 and 7 could not. The corresponding constrained clusterings defined on the collapsed haplotype tree are slightly less intuitive as repeated observations of the same haplotype (node) can belong to different clusters. In the collapsed haplotype network shown in Figure~\ref{haplotreesubdivided}, all clustered nodes must be directly connected within their cluster.  For simplicity, we shall refer to haplotype 4/6 as haplotype 4 from now on.

Formally, conditionally on a haplotype tree, the clustering model is defined as follows. We denote the set  of distinct haplotypes in the sequence set $\mathcal{S}$ (of size $N$) as $\mathcal{H}=\{H_1,\ldots,H_{n}\}$ with size $n$, and use $|H_i|$ to denote the number of copies of haplotype $H_i$ observed in the data.
Let $K$ denote the number of migrations, which is itself allowed to vary. Each migration event is associated with a  haplotype which migrated, denoted as $\mathbf{m}=\left\{m_1,\ldots,m_K\right\}$. Although colonisation events happen in order, here we do not model the events temporally, so the order of $\mathbf{m}$ is irrelevant. 
Note that the haplotypes in this list need not be distinct, as two different copies of the same haplotype may have colonised, or a single sequence may have colonised twice. The set of colonies/clusters with which each migrating haplotype is associated is denoted by $\mathcal{C}(m_k),\;k=1,\ldots,K$; in the example above, $K=3$, $\mathbf{m}=\left\{4,4,4\right\}$ and $\mathcal{C}(4)=\left\{\textrm{blue,yellow,pink,green}\right\}$, since all migrations were of the same haplotype. 
This means that, in general, the sample space of $\mathbf{m}$ has size ${n}^K/K!$.  

Conditionally on a set of migrating haplotypes, the space of constrained clusterings is then such that all observations of that haplotype must belong to one of the corresponding clusters $\mathcal{C}(m_k)$ (i.e.,~either the original cluster or the one which was a result of migration). Equivalently, all adjacent haplotypes must also belong to one of these clusters (unless one of them also migrated, and so on).

Once the clustering has been established, the geographical and ecological observations $Y_i,\;\;i=1,\ldots,N$ in each cluster $c_i$ are Normally distributed with mean $\mu_c$ and variance $\Sigma_c$, such that 
\begin{eqnarray}
Y_i&\sim& N(\mu_{c_i},\Sigma_{c_i}),\;\;i=1,\ldots,N,
\end{eqnarray}
where $c_i$ denotes the cluster of observation $i$. 

To complete the model, prior distributions are defined on the model parameters. The number of migrations is assumed to be uniform between 0 and $K_{\max}$ (corresponding to $1$ and $K_{\max}+1$ clusters). Other prior distributions (e.g.,~Poisson) could be used instead, but we do not explore this direction here.  The $|m_k|$ observations of each of the migrating haplotypes $m_k$ are each assigned uniformly to one of the clusters in $\mathcal{C}_k$, similarly with the $deg(m_k)$ clades connected to it (where degree represents the number of edges connected to node $m_k$), so the prior probability of each clustering conditional of the migrating haplotypes (and their clusters) is simply a combinatorial coefficient. 
\begin{equation}\label{colpriors}
\begin{split}
K \, \sim \, &\mathcal{U}\{0,\ldots,K_{\max}\},\\
\mathbf{m}\mid \mathcal{S}\,\sim\,& \textrm{Multinomial}\{|H_1|,\ldots,|H_n|\}\\
\textrm{and }p(\mathbf{c}\mid \mathbf{m},T) = &\prod_{k=1}^K{\left(\frac{1}{|\mathcal{C}_k|}\right)}^{|m_k|+deg(m_k)}
\end{split}
\end{equation}

The means and variances of each clustering are assigned different priors for the longitude-latitude versus the remaining covariates:
\begin{equation}
\begin{split}
\Sigma_{k,(1:2,1:2)}\,\sim \,&\mathcal{IW}(\gamma,\psi \mathbb{I}_2),\;k=1,\ldots,(K_{\max}+1),\\
\Sigma_{k,(3:d)}\,\sim \,&\mathcal{IG}(\gamma,\psi),\;k=1,\ldots,(K_{\max}+1),\\
\gamma\,\sim\, & \mathcal{U}\{4,\ldots,g\},\\
\mu_k\,|\,\Sigma_k\,\sim\,&\mathcal{N}\left(\mathbf{0},V\right),\;k=1,\ldots,(K_{\max}+1),
\end{split}
\end{equation}
and we assume that any off-diagonal entries of $\Sigma_k$  in dimensions $3:d$ are 0. 
By convention, the first two coordinates of $Y$ always represent  longitude and latitude, normalised such that the mean of both is zero and the average (between  longitude and latitude) variance 1, using the same normalising factor for both longitude and latitude to reflect the isotropy of the two dimensions. Note that longitude and latitude are treated as Euclidean coordinates, which means that datasets spanning a very large region may result in distorted results. 
The remaining coordinates  correspond to environmental or phenotypic characteristics (if available), which are normalised to sample mean 0 and marginal variance 1. We impose uncorrelated environmental/phenotypic characteristics by forcing the covariance matrices to be 0 on any off-diagonal entries except for the one correponding to longitude-latitude.  
This is because the concentration parameter $\gamma$ of an Inverse Wishart needs to be at least as $d_{\Sigma}+2$ in order to be well-defined, where $d_{\Sigma}$ is the dimension of the covariance matrix modelled. In our case, if we model the entire covariance matrix through an Inverse Wishart, $\gamma$ would be forced to a minimum of $3+d$, which (for moderate $d$) corresponds to low prior variance  and can be too restrictive. We thus restrict the Inverse-Wishart prior for the geographical covariates only and place independent Inverse-Gamma priors on the remaining diagonal elements of $\Sigma_k$.

Perhaps the most important prior distributions here are the ones relating to the shape $\Sigma$ of each cluster, namely the parameters of the Inverse-Wishart prior $\gamma$ and $\psi$, as these define the prior belief of the spread of each cluster. Although the parameter $\gamma$ is allowed to vary and hence can adapt depending on information from the data, nevertheless too large or too small values of $\psi$ (corresponding to a prior belief of geographically widely spread versus tiny clusters) will have an impact on the posterior inference. The default setting in \pkg{BPEC} is that clusters are a priori expected to span about $30\%$ of the total range. 

\section{Bayesian computation }\label{sec:bayes}
The entire model consists of the model of the root and tree posterior distribution together with the distribution of the migration and clustering model. Inferences are drawn simultaneously, such that we can borrow information from the tree to the migration parameters and vice versa. 
The complexity of this phylogeographic model implies that drawing inferences about the posterior distribution of the parameters is challenging. We proceed via tailored Markov chain Monte Carlo (MCMC) using a combination of adaptive proposals, auxiliary variables and data-driven proposals. This is especially crucial for the clustering, which here is restricted to tree-based clusterings, since the space of clusterings is vast and discrete without natural local moves. 

\subsection{Markov chain Monte Carlo sampler}
The Markov chain Monte Carlo sampler alternates between updates of the tree parameters and the clustering parameters. We adopt a scheme whereby updates of parameters are performed at varying frequencies, reflecting the difficulty of accepting or rejecting a move and allowing both local and global exploration of the parameter space. Four different updates are described below, which are then combined into a sampler at varying frequencies. 

The tree $T$, root $r$, colonised haplotypes $\mathbf{m}$, clustering $\mathbf{c}$ and cluster means $\boldsymbol{\mu}$ and variances $\boldsymbol{\Sigma}$.
\begin{itemize}
\item[1.] Conditionally on a given tree $T$, propose to change the root along with a mutation history. Accept or reject the proposed root and mutation history. 
\item[2a.] Conditionally on the root $r$, propose a new tree $T$ and mutation history uniformly. 
\item[2b.] Conditionally on the proposed tree $T$ propose to change one of the colonised haplotypes in $\mathbf{m}$.
\item[2c.] Conditionally on the colonised haplotypes, propose to change the set of clusterings $\mathbf{c}$ along with the means $\boldsymbol{\mu}$ and variances $\boldsymbol{\Sigma}$ of each cluster. 
\item[2d.] The proposed tree topology and history, root, clustering and means and variances are accepted or rejected together. However, steps (2a), (2b) and (2c) need not all occur at the same time. Specifically, steps (2a-b) are only performed (roughly) every 5th iteration. 
\item[3.] Conditionally on a given clustering, update the cluster means conditionally on all other parameters, and subsequently the sample covariance conditionally on all other parameters. 
\item[4.] Propose to increase or decrease the number of clusters. Then propose to add or subtract a colonised sequence, then set of clusterings together with means and variances of each cluster. Accept or reject the entire move. 

\end{itemize}
The precise mechanics of the sampler are not shown here; some additional technical issues are discussed in Appendix~\ref{appendix:issues}.

\subsection{Technical considerations}
Almost as important as how the method works is when it is sound to use (or not). Since the package is intended to be used primarily by practitioners, one of the aims of this paper is to clarify what types questions \pkg{BPEC} can potentially answer as well as what underlying assumptions are necessary and implicit.
					
Bayesian Phylogeographic and Ecological Clustering assumes that non-recombinant (typically mtDNA) data are available from a set of geographical locations (in the form of longitude/latitude). The haplotype tree model takes a relaxed parsimony approach which may be unreliable under conditions of mutational saturation or excessive homoplasy. \pkg{BPEC} is programmed to produce appropriate error messages to inform the user in such cases, but will not be foolproof.
					
The geographical model assumes a constant population size and migration rate, and thus as real data departs from this model the inferences from \pkg{BPEC} are expected to depart from the true demographic history. However, simulation analyses will be required to address this quantitatively. Also, the clustering and migration model does not explicitly take into account geographical distance between clusters. It simply separates observations in distinct geographical clusters. Therefore, it is possible for a migration to result in two distant clusters.
					
Notice that we assume a uniform prior over the number of migrations $K$. In general, $K$ migra- tions can lead to up to $K + 1$ clusters; often, however, some of these may be empty, resulting in fewer `effective' migrations. The uniform prior applies to the total number of migrations rather than the number of effective ones, whereas the posterior distribution over the number of migrations actually refers to effective migrations. This somewhat convoluted approach is preferred because enumerating scenarios of different effective migrations is computationally cumbersome.
					
As discussed earlier, an important consideration when using \pkg{BPEC} for the inference of ancestral areas is the distribution of haplotype observations within each location. Since ancestral area probabilities are determined through the proportion of inferred ancestral haplotypes, a site with, for example, a single haplotype which happens to be ancestral, will always result in high probability of being ancestral. Consequently, ancestral location probabilities should be more reliable when there are more observations per location. It also frequently occurs that uncertainty about the root haplotype is high, where a range of different haplotypes carry significant posterior mass. As long as no convergence errors are reported, this is not a convergence issue but merely reflects uncertainty in the data.
					
One of the limitations of Markov chain Monte Carlo methods is that the samplers require a large number of iterations to satisfy convergence diagnostics. The convergence diagnostics in \pkg{BPEC} are split into two pieces: convergence of the clustering and convergence of the root haplotype. If either of these two pieces has not converged, the sampler will return an error to that effect. Ideally, both pieces should satisfy the convergence diagnostics; however, it is sometimes the case (especially when dealing with a large number of clusters) that, for any reasonable number of MCMC iterations, the diagnostics fail. In these cases, inferences should be taken with caution.
					
\pkg{BPEC} cannot deal with unknown nucleotides and will ignore any nucleotide sites at which at least one of the sequences has an ambiguity code. This means that ambiguous nucleotides result in information loss. On the other hand, \pkg{BPEC} will treat true alignment gaps `-' as a 5th character such that a deletion/insertion is treated as a type of mutation. Care should be taken in the interpretation of the output when lots of missing nucleotides are present, since this could lead to significant loss of resolution \citep[][]{joly2007haplotype}.

\section{Brown frog data}\label{sec:brownfrog}
\label{brownfrogs}
The \pkg{BPEC} package will be implemented on a brown frog dataset which will be used throughout the next few sections for illustration. 
We used 40 mitochondrial cytochrome b sequences of Near Eastern brown frogs - \emph{Rana macrocnemis} (Boulenger, 1885) to demonstrate a combined phylogeographic and ecological analysis with \pkg{BPEC}. Previous molecular analyses \citep{tarkhnishvili2001humid,veith2003palaeoclimatic,veith2003climatic} have attributed range expansion and fragmentation triggered by Pleistocene glaciation cycles as drivers of demographic change within the brown frog.

\begin{figure}[h!]
\centering
\includegraphics{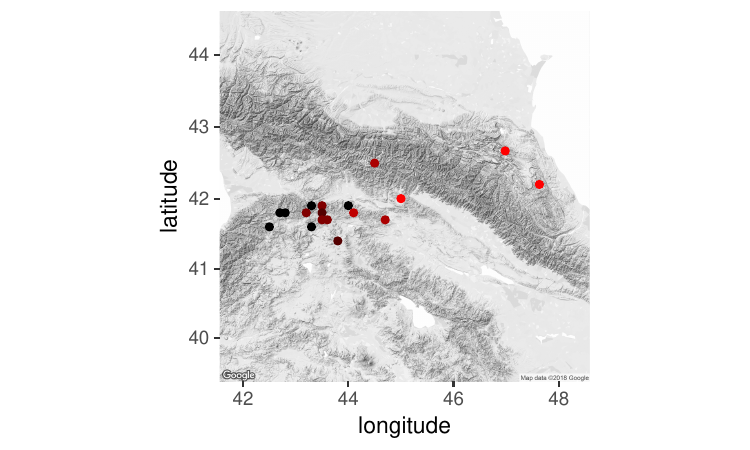}
\caption{\label{BrownFrogMap} The sampling localities for \emph{R.~m.~camerani} and \emph{R.~m.~macrocnemis} overlayed onto the map. Colour represents proportion of \emph{R.~m.~camerani} versus \emph{R.~m.~macrocnemis} individuals sampled, with red corresponding to 100\% \emph{R.~m.~camerani}, black 100\% \emph{R.~m.~macrocnemis} and brown corresponding to populations with both \emph{R.~m.~camerani} and \emph{R.~m.~macrocnemis}  present.}
\end{figure}

\emph{R.~macrocnemis} is represented by a number of recognised subspecies across its entire range, and here we focus on two widespread subspecies that are geographically distinct in the south-west Caucasus and separated by a narrow transition zone \citep{tarkhnishvili2001humid}. The nominotypic \emph{R.~macrocnemis macrocnemis} (Boulenger, 1885) is found on the forested slopes of the Trialeti ridge northwest and in montane meadows on both sides of the Great Caucasus, while \emph{R.~macrocnemis camerani} (Boulenger, 1885)  occurs in southern Georgia on the Javakheti plateau \citep{tarkhnishvili2001humid}. A map of the sampling localities, indicating proportion of \emph{R.~macrocnemis macrocnemis} versus \emph{R.~macrocnemis camerani}, is shown in Figure~\ref{BrownFrogMap}.

\pkg{BPEC} was applied to investigate geographic and environmental aggregation of haplotypes within \emph{R.~macrocnemis}. We included predictive environmental and climate covariates (topographic and land cover conditions, and annual trend patterns of temperature, precipitation and seasonality) to examine environment and geography as agents for the structuring of genetic variation.  Grid-based attribute values of a set of predictor 
variables associated with each cell position of the map layers 
were subsequently extracted at the point locations of the 
georeferenced mtDNA haplotypes from six raster grids by means of 
the \code{extract} function of the \pkg{raster} package \citep{hijmans2015raster}: 
four bioclimatic variables (Annual Mean 
Temperature (degrees Celsius x 10), Temperature Annual 
Range (100 $\times$ standard deviation of monthly mean temperature), 
Annual Precipitation (in mm),  Precipitation 
Seasonality (Coefficient of Variation (CV)), from the bioclim
database available in the \pkg{dismo} package \citep{hijmans2014species},
altitude in meters as a proxy for a digital elevation model and the land cover
map (GLC2000) from the subdomain land cover/land use housed under
\url{http://worldgrids.org/} global environmental layers.
We re-classified the total information of the land cover map into 
two classes of forested and non-forested areas to introduce a 
simplistic landscape dependent habitat variable (COV). These six 
variables altogether describe climatic, topographic, and land 
cover conditions that are potentially informative predictors in terms of species distribution. 

\section{User interface}\label{sec:interface}

\subsection{Inputs}
\pkg{BPEC} takes two main inputs: the set of mtDNA sequences (in NEXUS format) and the set of coordinates and haplotypes observed in each location. Sequences need not be collapsed into unique haplotypes, but labelling of sequences in the NEXUS file and the locations file must be consistent. In order to load these two variables into \proglang{R} from two files called \code{haplotypes.nex} and \code{coordsLocsFile.txt} (for example), the following commands can be used. For an example of input files, use the files provided through \code{system.file("haplotypes.nex",package = "BPEC")} and \code{system.file("coordsLocsFile.txt",package = "BPEC")} or see Supplementary materials of the manuscript.

The sequences can be loaded using the \code{bpec.loadSeq} command.

\begin{Schunk}
\begin{Sinput}
R> library("BPEC")
R> rawSeqs <- bpec.loadSeq("haplotypes.nex")
\end{Sinput}
\end{Schunk}

The file \code{coordsLocsFile.txt}, containing the list of coordinates, covariates and haplotypes, needs to have (in each row): latitude, longitude, environmental/phenotypic covariate values (if available), plus a set of numbers corresponding to the haplotypes/sequences with these attributes. For example, for two locations with 5 observations in total from 3 haplotypes, with no additional covariates, the file might read 
\begin{Code}
40.3 45.2    1    2    2
45.3 50.1    2    3
\end{Code}

All haplotypes/sequences found in a location can be entered in one line, or only one per row, such that we could also have used
\begin{Code}
40.3 45.2    1    2
40.3 45.2    2
45.3 50.1    2    3
\end{Code}
or any such combination.  

When additional environmental or phenotypic covariates are available, these can also be entered as a column right after the longitude and latitude, such as 
\begin{Code}
40.3 45.2 18.1    1 
40.3 45.2 22.5    2    2
45.3 50.1 25.0    2    3
\end{Code}
where $18.1,22.5,25.0$ are, for example, temperatures. Names for the covariates at each location can optionally be provided through the header row using the option \code{header = TRUE}, these will later appear in the output plots to aid interpretation.
\begin{Code}
lon  lat  temp  
40.3 45.2 18.1    1 
40.3 45.2 22.5    2    2
45.3 50.1 25.0    2    3
\end{Code}
Environmental covariates can be extracted, for example, from publicly available databases such as bioclim by means of the \proglang{R} package \pkg{raster} \citep{hijmans2015raster}.

In order to load the file containing the coordinates, covariates and observed haplotypes/sequences of each location, use the \code{bpec.loadCoords} command below. Use the option \code{header = TRUE} when the first row includes variable names. 
\begin{Schunk}
\begin{Sinput}
R> coordsLocs <- bpec.loadCoords("coordsLocsFile.txt", header = TRUE)
\end{Sinput}
\end{Schunk}

The brown frog dataset is in-built and can be loaded through
\begin{Schunk}
\begin{Sinput}
R> data("MacrocnemisRawSeqs")
R> data("MacrocnemisCoordsLocs")
R> rawSeqs <- MacrocnemisRawSeqs
R> coordsLocs <- MacrocnemisCoordsLocs
\end{Sinput}
\end{Schunk}
which contain the 40 sequences together with their corresponding longitude/latitude, along with 6 environmental covariates.  Other datasets that are available in \pkg{BPEC} can be found using \code{data(package = "BPEC")}.

\subsection{Main MCMC command and options}
Once the \code{rawSeqs} and \code{coordsLocs} variables have been loaded, the Markov chain Monte Carlo sampler can be run through the command

\begin{Schunk}
\begin{Sinput}
R> bpecout <- bpec.mcmc(rawSeqs, coordsLocs, maxMig = 3, iter = 1000000, 
+                      ds = 3, postSamples = 1000, dims = 8)
\end{Sinput}
\end{Schunk}

The arguments are described in Table~\ref{tab:bpecArguments}.
\begin{table}
\begin{tabular}[h]{r p{13cm}}
\hline
\hline
\\[-1em]
\multicolumn{2}{l}{\hphantom{blablablablab}\code{bpec.mcmc()}}\\
\\[-1em]
\\[-1em]
\emph{Argument} & \emph{Description}\\
\hline

\code{maxMig} & the maximum number of migrations to be considered. In terms of inference, the higher \code{maxMig}, the better the results, since more models are considered. However, that comes at a computational cost. We recommend using a low but intuitive value based on the study system to begin an iterative assessment. For example, if using a value of 6 (corresponding to 7 clusters), and if the inference shows significant posterior probability on 7 clusters, increase \code{maxMig} and re-run.  Similarly, if e.g.,~a value of 5 is used and convergence diagnostics are not satisfied, but posterior mass seems to be minimal around 4/5 migrations, then one can reduce \code{maxMig} to 4 (which will reduce complexity) and re-run. \vspace{3mm}\\

\code{iter} & the number of MCMC iterations to run the sampler for. By default, two chains will be run from different starting values. The value of \code{iter} is important, as it will determine how long the chains will run for and whether convergence (both in terms of the root haplotype as well as the clustering) diagnostics will be satisfied. A value of 100,000 is usually reasonable to start with; if convergence diagnostics are not satisfied, or if the post-processing plots look inconsistent, increase \code{iter} by a factor of 10 (and so on). \vspace{3mm}\\

\code{ds} & the parsimony relaxation parameter $d_s$. We recommend starting with $d_s=0$ and increasing once reasonable values of \code{iter} and \code{maxMig} have been established. Note that increasing $d_s$ past an (unknown) value $d_{max}$, which depends on the individual dataset, has no effect on the inference. \vspace{3mm}\\

\code{postSamples} & the number of posterior samples (per chain) to be saved for posterior summary statistics. We recommend using a value around 1,000. The higher the better for inference, but this comes at a memory storage cost. \vspace{3mm}\\

\code{dims} & the number of covariates (including longitude and latitude) available. If only geographical data are used (and no environmental or phenotypic information), \code{dims = 2}. Otherwise increase as appropriate. \vspace{2mm}\\
\hline
\end{tabular}
\caption{\label{tab:bpecArguments} The list of inputs required to use \code{bpec.mcmc()}.}
\end{table}

In the case of the brown frog dataset, the dimensionality of the data was \code{dims = 8} (geographical dimensions longitude and latitude plus the six additional environmental covariates). We ran the \pkg{BPEC} analysis taking the maximum parsimony level option at \code{ds = 0}, increasing up to \code{ds = 3} (to potentially explore more candidate trees) for 1,000,000 iterations (\code{iter}) each. No change to the results was observed, since the brown frog haplotypes formed a fully connected tree without missing intermediate haplotypes. Convergence diagnostics of the maximum a posteriori clusterings and root were not violated (i.e.,~no convergence error message was reported). The output of the function is shown below. 
\begin{Schunk}
\begin{Soutput}
Starting bpec...
Inferring possible missing sequences....
Counting loops in the network...

The program found no loops that need to be resolved in the network

Number of iterations is 1000000
Number of saved iterations 1000
Sample size is 40
Effective sequence length is 8
Total number of haplotypes (including missing) 10
Dimension is 8
Parsimony relaxation is 3
Maximum number of migrations is 3

Starting MCMC sampler (burn-in ends at 90% and acceptance rate re-started):
Chain 1: |====================|100% (accepted samples 2763 time 24 minutes)
Chain 2: |====================|100% (accepted samples 2823 time 49 minutes)

The most likely root node is 2
The most likely ancestral locations are 38,34,4
\end{Soutput}
\end{Schunk}

\subsection{Outputs}
The \code{bpec.mcmc} command outputs an \proglang{R} object of class \code{BPEC} which can be summarised using generic functions such as \code{plot()}, \code{summary()} and \code{plot()}, as well as accessor functions \code{input()}, \code{preproc()}, \code{output.tree()}, \code{output.clust()}, \code{output.mcmc()}. The output of each of these accessor functions is shown in Tables~\ref{tab:input}-\ref{tab:mcmc}.

\begin{table}
\begin{tabular}[h]{r p{13cm}}
\hline
\hline
\\[-1em]
\multicolumn{2}{l}{\hphantom{blablablablab}\code{input()}}\\
\\[-1em]
\\[-1em]
\emph{Output} & \emph{Description}\\
\hline
\vspace{2mm}\code{seqCountOrig}& the number of sequences in the data.\\
\code{seqLengthOrig}& the length of the input sequences.\\
\code{iter}& the number of MCMC iterations.\\
\code{ds}& the parsimony relaxation parameter.\\
\code{coordsLocs}& the input coordinates (and optional additional ecological measurements) and their corresponding sequence indices. \\
\code{coordsDims}& the dimension of the input measurements (2 if purely longitude and latitude, +1 for every additional one).\\
\code{locNo}& the number of distinct sampling locations. \\
\code{locData}& the coordinates and measurements of each sampled sequence. \vspace{2mm}\\
\hline
\hline
\end{tabular}
\caption{\label{tab:input}The list of outputs of \code{input()}, corresponding to all the inputs and arguments that were provided to \code{bpec.mcmc()}.}
\end{table}

\begin{table}
\begin{tabular}[h]{r p{13cm}}
\hline
\hline
\\[-1em]
\multicolumn{2}{l}{\hphantom{blablablablab}\code{preproc()}}\\
\\[-1em]
%\hline 
\\[-1em]
\emph{Output} & \emph{Description}\\
\hline
\code{seq}& The output DNA sequences of distinct haplotypes, collapsed to effective nucleotide sites (both sampled and missing sequences which were inferred).\\
\code{seqsFile}& A vector of the numerical labels of each haplotype.\\
\code{seqLabels}&  Correspondence vector for each of the processed observations to the original haplotype labels.\\
\code{seqIndices}& Correspondence vector for each of the original observations to the resulting haplotype labels.\\
\code{seqLength}& The effective length of the input sequences, given by the number of variable nucleotide sites which are informative. In other words, if two or more nucleotide sites describe the same subsets of sequences, then they are collapsed to a single informative nucleotide.\\
\code{noSamples}&  The number of times each haplotype was observed in the sample. \\
\code{count}&  The number of output sequences.\vspace{2mm}\\
\hline
\hline
\end{tabular}
\caption{\label{tab:data}The list of outputs of \code{preproc()}, corresponding to values arising from the data before the Bayesian analysis.}
\end{table}

\begin{table}
\begin{tabular}[h]{r p{13cm}}
\hline
\hline
\\[-1em]
\multicolumn{2}{l}{\hphantom{blablablablab}\code{output.tree()}}\\
\\[-1em]
%\hline 
\\[-1em]
\emph{Output} & \emph{Description}\\
\hline
 \code{clado}& the adjacency matrix for the maximum a posteriori tree in vectorised format. For two haplotypes \code{i,j}, the \code{(i,j)}th entry of the matrix is 1 if the haplotypes are connected in the network and 0 otherwise.\\
 \code{levels}& Starting from the root (level 0) all the way to the tips, the discrete depth for the maximum a posteriori tree.\\
 \code{edgeTotalProb}& Posterior probabilities of each edge being present in the tree, so that any edge which is not part of a loop will have posterior probability 1.\\
 \code{rootProbs}& a vector of the posterior probabilities that each haplotype is the root of the tree.\\
 \code{treeEdges}& contains the same information as \code{cladoR}, but in a different format.  The set of edges (from and to haplotypes) of the maximum a posteriori haplotype tree are represented as an edge list of from/to vectors which could be used in the graph and network modelling \proglang{R} package \pkg{igraph} \citep{Csardi:2006uq} if needed.\\
 \code{rootLocProbs}& a vector of the posterior probabilities of each sampling location being the most ancestral location. If several rows in the file \code{coordsLocsFile.txt} correspond to the same geographical location, the first of these will carry the total posterior probability for the location, with the remaining having 0.  \\
\code{migProbs}& a vector of the posterior probabilities of $\left\{\texttt{0}\ldots\texttt{maxMig}\right\}$ migrations.\vspace{2mm}\\
\hline
\hline
\end{tabular}
\caption{\label{tab:tree}The list of outputs of \code{output.tree()}, corresponding to the output of the tree model.}
\end{table}

\begin{table}
\begin{tabular}[h]{r p{13cm}}
\hline
\hline
\\[-1em]
\multicolumn{2}{l}{\hphantom{blablablablab}\code{output.clust()}}\\
\\[-1em]
%\hline 
\\[-1em]
\emph{Output} & \emph{Description}\\
\hline
 \code{sampleMeans}& a set of \code{postSamples} posterior samples of the cluster centres.\\
 \code{sampleCovs}& a set of \code{postSamples} posterior samples of the cluster covariances.\\
 \code{sampleIndices}& a set of posterior samples of the cluster allocations of each observation.\\
 \code{clusterProbs}& for each haplotype, posterior probabilities that it belongs to each cluster.\vspace{2mm}\\
\hline
\hline
\end{tabular}
\caption{\label{tab:clust}The list of outputs of \code{output.clust()}, corresponding to the output of the geographical clustering  model.}
\end{table}

\begin{table}
\begin{tabular}[h]{r p{13cm}}
\hline
\hline
\\[-1em]
\multicolumn{2}{l}{\hphantom{blablablablab}\code{output.mcmc()}}\\
\\[-1em]
%\hline 
\\[-1em]
\emph{Output} & \emph{Description}\\
\hline
 \code{MCMCparams}& various tuning parameters used in the MCMC sampler, this is only important for development.\\
 \code{codaInput}& Posterior samples from the two MCMC chains for the cluster means, cluster covariance entries, as well as the root haplotype. Note that, since the number of clusters varies from iteration to iteration, some samples are simply draws from the prior (corresponding to empty clusters). This variable can be loaded directly into the \pkg{coda} package \citep{plummer2006coda} for convergence analysis.\vspace{2mm}\\
\hline
\hline
\end{tabular}
\caption{\label{tab:mcmc}The list of outputs of \code{output.mcmc()}, corresponding to technical MCMC aspects.}
\end{table}

\subsection{Visualizations and post-processing}
As described in the previous section, the Markov chain Monte Carlo sampler returns many different types of outputs. In order to obtain a summarised picture of the inference, a number of visualizations are available through \pkg{BPEC} to aid interpretation.

\subsubsection{Geographical contour plot}
The command \code{bpec.contourPlot} provides a colour-coded contour plot of the geographical clusters superimposed onto a map (provided accurate longitude and latitude coordinates have been provided) using
\begin{Schunk}
\begin{Sinput}
R> par(mar = c(0, 0, 0, 0))
R> bpec.contourPlot(bpecout, GoogleEarth = 0, mapType = 'google', 
+                   colorCode = c(7, 5, 6, 3, 2), mapCentre = NULL, zoom = 7)
\end{Sinput}
\end{Schunk}
In order to convey not only posterior means but also uncertainty, a set of posterior draws of these contours are plotted using transparency, so that the user can assess the stability of the inference. 

The sampling locations are also shown on this contour plot, with the top three sampling locations in terms of their probability of being ancestral shown as larger points. The precise posterior probabilities (which may all be low in the presence of uncertainty) of each of the localities being ancestral can be found through \code{output.tree(bpecout)$rootLocProbsR}. 

 The colours can be changed through the optional argument \code{colorCode} (with default value \code{(7,5,6,3,2,8,4,9)}) which controls the colour of the first, second, third cluster etc; if not specified, the default colour scheme is used.  There are four options for the argument \code{mapType}: \code{`none'} will show the posterior distribution of the clusters against a white background, \code{`plain'} will use the in-built outline \proglang{R} maps, \code{`google'} will superimpose the contours on a map downloaded from Google maps (requires an internet connection), and \code{`osm'} will do the same using OpenStreetMap. The optional arguments \code{mapCentre} and \code{zoom} allow the user to specify the centre of the map and level of zooming when using the Google maps option. 
 
\begin{figure}[h!]
\centering
\includegraphics{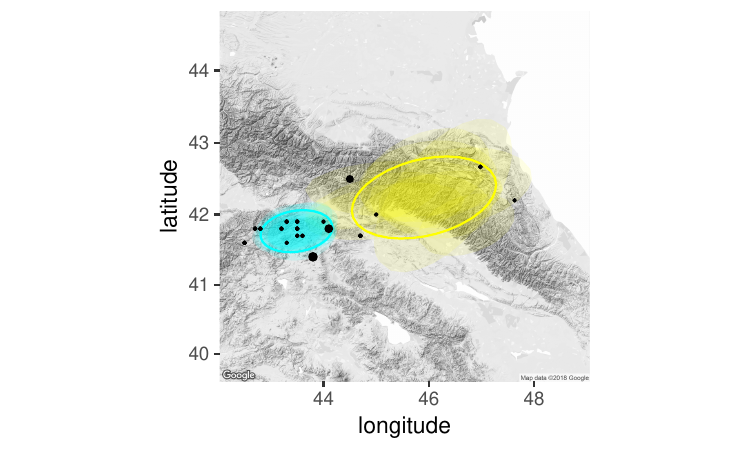}
\caption{\label{ContourPlot} An example of the contour plot for the brown frog dataset using \code{bpec.contourPlot}. Each transparent geographical ellipse represents a posterior draw for the geographic centre of the cluster within the 50\% level contour of that draw. The 50\% contour represents the boundary where probability density of the cluster is 50\% of the maximum density (i.e.,~the centre of the cluster). Solid ellipses represent posterior means. Larger triangles represent most likely ancestral locations. The black jagged lines show the outline of the geographical map of the area.
}
\end{figure}

 In the case of the brown frog dataset, the contour plot is shown in Figure~\ref{ContourPlot}.
 The posterior mass for the number of clusters strongly concentrates around 2 (as indicated by the output \code{output.clust(bpecout)$migProbs}), with the posterior probability of 2 clusters being greater than 0.99. The yellow cluster can be taxonomically aligned to the subspecies \emph{R.~m.~macrocnemis} lineage, while the turquoise cluster includes individuals of \emph{R.~m.~macrocnemis} from the humid and forested mountain region, and individuals assigned to \emph{R.~m.~camerani} from the drier area of the southern treeless mountain steppe habitats of the Javakheti plateau. The contour ellipses overlap in the heart of the geographic transition zone south of the Minor Caucasus. 

Instead of using the \proglang{R} interface, the contour plot can also be exported into Google Earth primary exchange format using the option \code{GoogleEarth = 1}. This will produce a set of files with extension \code{kml} which can be loaded directly into Google Earth.

\begin{figure}[h!]
\centering
\includegraphics[width=10cm]{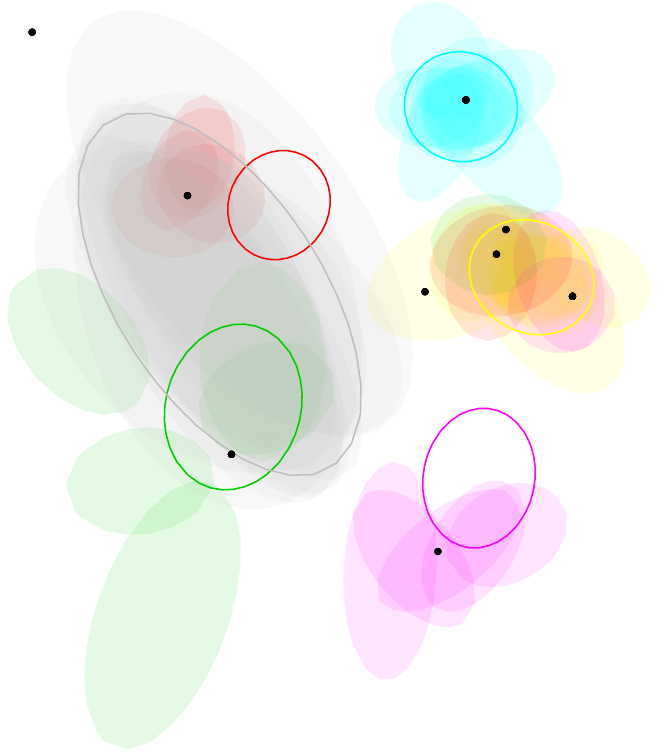}
\caption{\label{messy} A toy example of a posterior distribution of phylogeographic clustering which shows high uncertainty. Many of the clusters `jump' from one location to another from iteration to iteration, indicating uncertainty about the location (and number) of clusters. }
\end{figure}

Finally, a `messy' looking plot such as the toy example in Figure~\ref{messy} either implies poor MCMC convergence or high uncertainty in terms of the clustering. 

\subsubsection{Environmental and/or phenotypic covariates plot}
In cases where environmental or phenotypic covariates have also been used, posterior draws for the distribution of the covariates within clusters are available through cluster means \code{output.clust(bpecout)$sampleMeans} and covariances \code{output.clust(bpecout)$sampleCovs}. These can be summarized through posterior medians and 5/95\% credible regions, colour-coded using the same coding as the contour plot. To aid plotting and interpretation,  the covariate names of each of the columns of \code{coordsLocs} are used. The first two (corresponding to longitude and latitude) are automatically ignored in this function. 
\begin{Schunk}
\begin{Sinput}
R> par(mfrow = c(2, 3))
R> bpec.covariatesPlot(bpecout, colorCode = c(7, 5, 6, 3, 2)) 
\end{Sinput}
\end{Schunk}

\begin{figure}[h!]
\centering
\includegraphics{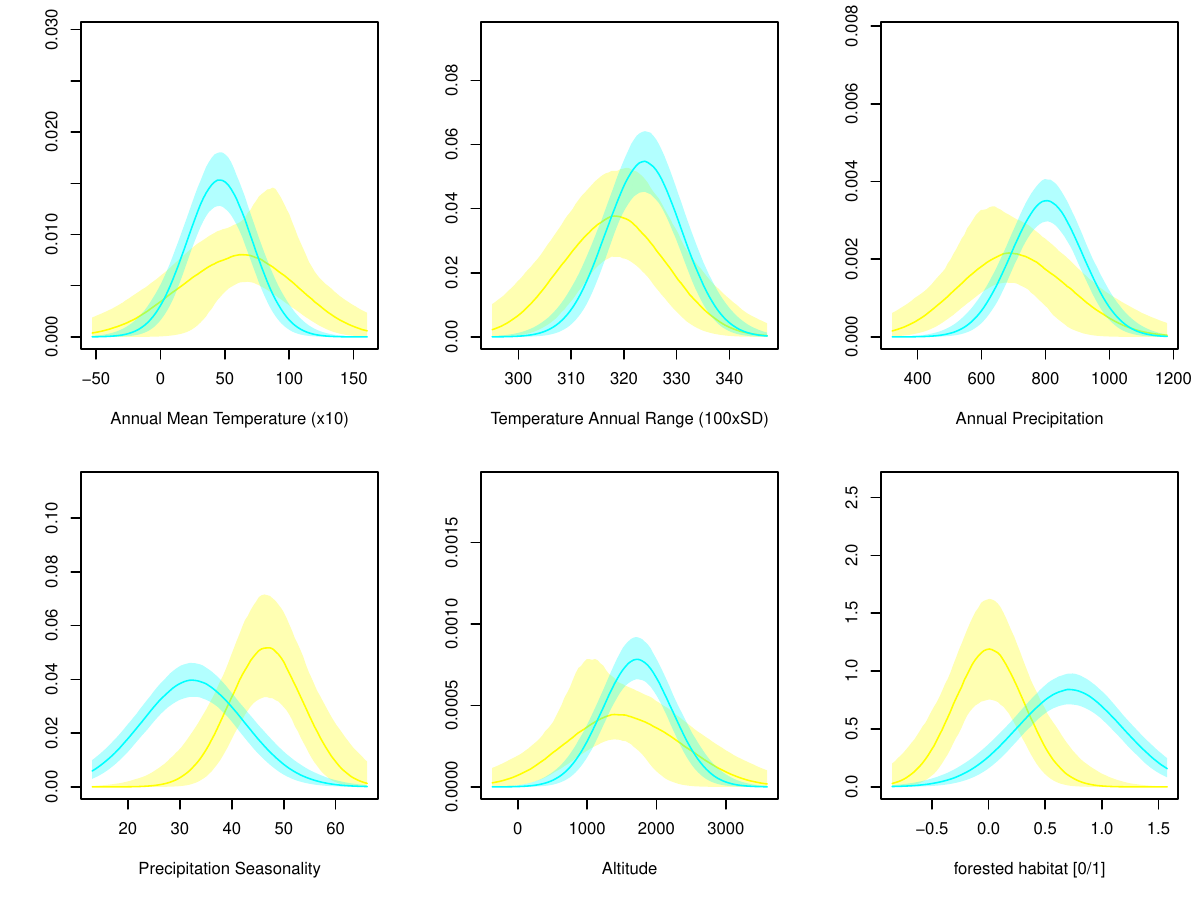}
\caption{\label{CovariatesPlot} Plot of the distribution of the covariates for each cluster for the brown frog dataset using \code{bpec.covariatesPlot}. Shaded regions correspond to 5\% and 95\% pointwise credibility bands of each cluster, with solid lines showing the pointwise median. Colour corresponds to the same clusters as the contour plot above. }
\end{figure}

The plot produced in the case of the brown frog dataset is shown in Figure~\ref{CovariatesPlot}.

\subsubsection{Clustered tree plot}
To visualize the maximum a posteriori haplotype tree, the command \code{bpec.treePlot} plots the haplotype tree most supported by the data.  The size of each node in the tree represents the number of times each haplotype was observed, black dots corresponding to missing intermediate haplotypes.  The thickness of each edge represents the posterior probability that each mutation occurred (thin edges corresponding to mutations with high uncertainty). Observed haplotypes are colour-coded according to their posterior probability of belonging to each cluster. As long as the same \code{colorCode} variable is used, the cluster colours correspond to the ones used in the geographical and covariate contour plots.
\begin{Schunk}
\begin{Sinput}
R> bpec.tree <- bpec.treePlot(bpecout, colorCode = c(7, 5, 6, 3, 2))   
\end{Sinput}
\end{Schunk}

\begin{figure}[h!]
\centering
\includegraphics{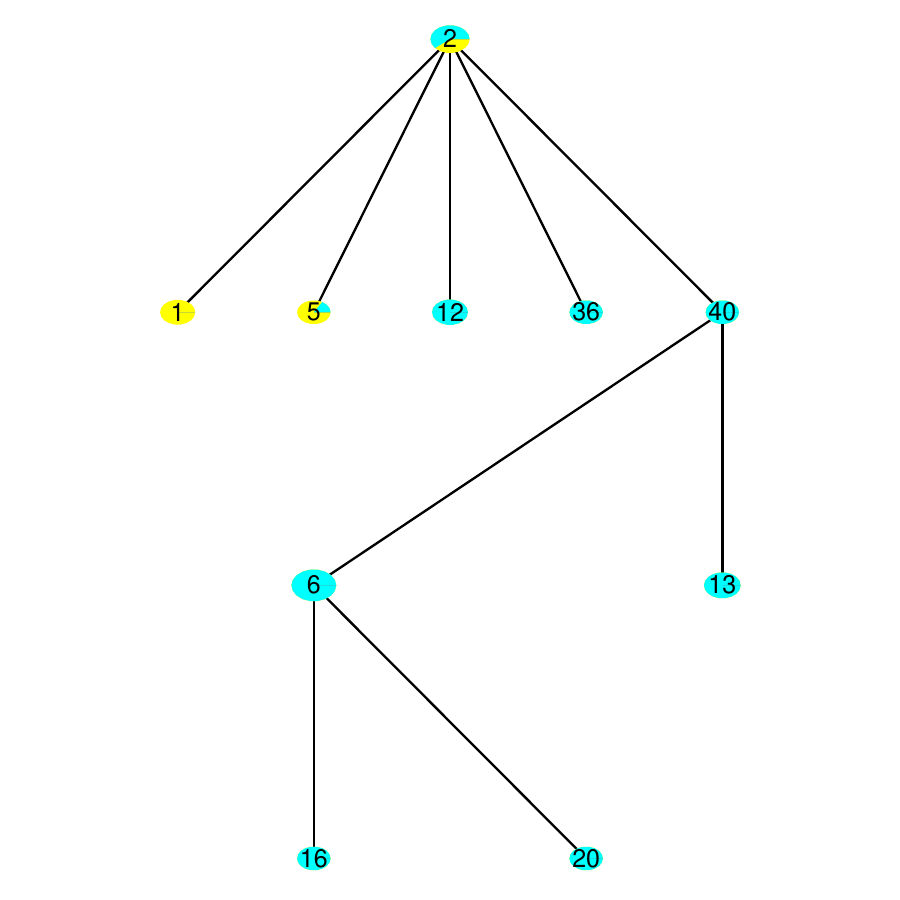}
\caption{\label{TreePlot} Clustered tree plot of the brown frog dataset using \code{bpec.treePlot}. Colour corresponds to cluster membership probability and size of node to the number of individuals sampled with each sequence. Edges represent single effective mutations and black dots represent unobserved intermediate haplotypes. In this case all edges have effectively no posterior uncertainty under the model, so they all appear with equal thickness.
}
\end{figure}

The corresponding plot for the brown frog dataset is shown in Figure~\ref{TreePlot}. \pkg{BPEC} collapsed sequences to 10 distinct haplotypes with effective length 8 displayed in the maximum a posteriori haplotype tree shown.

\subsubsection{Tree plot on geographical map}
The tree plot can also be partially visualised geographically through the \code{bpec.geoTree} command which superimposes the haplotype tree onto a map through a file that can be loaded into Google Earth. The function uses the \pkg{igraph} package \citep{Csardi:2006uq} as well as \pkg{phytools} \citep{Revell:2012fk,valiente2010combinatorial} in order to visualise the network as an interactive tree. Finally, to overlay the tree onto the map, the archived library \pkg{R2G2} is used \citep{arrigo2012quantitative,arrigo2013R2G2}.
Since haplotypes can be observed in multiple locations, clicking on particular nodes of the tree shows the locations where each copy of the haplotype was found. However, when multiple haplotypes were found in a single location, only one will be displayed, so the \code{bpec.geoTree} may not tell the whole story. Also note that only existing tip haplotypes are possible to identify on the map. 

\begin{Schunk}
\begin{Sinput}
R> bpec.geo <- bpec.geoTree(bpecout, file = "GoogleEarthTree.kml")
\end{Sinput}
\end{Schunk}

Tip haplotypes are connected to a tree by a single branch, internal node haplotypes have three or more connections, whereas branch haplotypes exactly two connections. 

\section{Analysis of the brown frog data}\label{sec:analysis}
In the case of the brown frog data, the three locations with highest probability of being ancestral are approximately located at the intersection between the yellow and turquoise cluster, shown as larger dots in Figure~\ref{ContourPlot}. These three locations correspond to (a) Paravani lake, treeless mountain steppe, 2100m, Javakheti plateau, posterior probability 24\%, (b) Tsalka, treeless mountain steppe, close to the southern slopes of the Trialeti Ridge, posterior probability 12\% and (c) Cross Mountain Pass, alpine habitat, 2000-2500m, Great Caucasus, posterior probability 10\%. However, it is important to condition any conclusions drawn from these inferences on their associated probability values, and in the case of \emph{R.~macrocnemis} it is clear that there is high uncertainty associated with these inferences that is related to the limited information content of the data rather than issues of convergence.  Further sampling (more localities and more individuals per locality) may improve posterior probabilities, but it may also be possible to develop \pkg{BPEC} to incorporate outgroup sequences for the inference of ancestral haplotypes (see Section~\ref{sec:discussion}), something that should improve posterior probabilities for ancestral areas.

Differences within bioclimatic (annual mean temperature, annual range of temperature, annual precipitation, precipitation seasonality) and altitude variables between the two clusters (Figure~\ref{CovariatesPlot}) are largely due to their mean values rather than their variances. Differences in means are especially apparent for the annual mean temperature (> 5\celsius ~for the yellow cluster  and around 5\celsius ~for the turquoise cluster), annual range of temperature (high amplitude of variation, typical for mountain climates: CV < 320\% for the yellow cluster, > 320\% for populations of the turquoise cluster), annual precipitation (higher for populations of the turquoise cluster, nearly 800mm), annual distribution of precipitation (much higher in the yellow cluster, CV > 45\% which results in higher variation in the timing and intensity of annual precipitation). Altitude is rather similar around 1500-1800m. Finally, the landscape dependent variable `open vs. forested habitat' is clearly different for the clusters. These findings suggest that individuals within the sampled area for \emph{R.~macrocnemis} are best described by two geographic clusters of mtDNA sequence variation and that they also differ with respect to specific environmental conditions. These data therefore offer support to the hypothesis that both processes of geographic isolation and divergent selection have contributed to diversification within the group, with the suggestion that taxonomy recognises these entities.  As such, the results of \pkg{BPEC} provide specific hypotheses than can be further tested with a more extensive genetic marker based approach for hypothesis testing (see Section~\ref{sec:discussion}) .

\section{Discussion}\label{sec:discussion}
We have described \pkg{BPEC}, an implementation of the phylogeographic and ecological clustering methods described in \cite{manolopoulou2011bayesian,manolopoulou2012phylogeographic}. We have introduced several visualization and post-processing tools in order to aid data analysis and interpretation, along with details of the significance of different types of output. \pkg{BPEC} will continue to be improved. The main focus of the extensions will revolve around speeding up the convergence of the sampler and improving the approximation stemming from the auxiliary tree parameter.

We recommend caution when extrapolating conclusions from \pkg{BPEC} output, and as is the case for many software packages it is important that users do not take a black box approach. Users should condition their conclusions on the biology of their organism of interest, the completeness of their sampling, and the idiosyncrasies of their data (e.g.~the proportion of unsampled haplotypes). In terms of extensions to the actual model, more generic evolutionary models for subdivided haplotype trees will be gradually introduced, such as the one recently developed  by \cite{de2015new}. Similarly, explicitly modelling the migration process as a spatial transition will allow additional information from the spatial distribution to inform the tree and vice versa. As currently configured, \pkg{BPEC} is best treated as a tool that can potentially reduce model space for subsequent hypothesis testing.  As an example, in the case of the brown frog \emph{Rana macrocnemis} \pkg{BPEC} identified geographic clusters of mtDNA sequence variation that are associated with differing environmental environmental conditions that could underpin divergent selection.  Thus, \pkg{BPEC} presents evidence that both neutral and selective processes are driving diversification within the group. However, as \pkg{BPEC} is limited to the analysis of a single DNA sequence locus, inferences should not be extrapolated to ultimate biological/ecological conclusions. \pkg{BPEC} should lend itself to the integration of inferences across multiple loci within a species, and this is an area that we are investigating for future updates.  Of particular relevance is the increasing accessibility of reduced genome sequencing data \citep{mccormack2013applications} that can provide up to tens of thousands of loci per individual.  Filtering for loci characterised by multiple SNPs could provide a rich data source for a multi-locus \pkg{BPEC} implementation.  Analogous to a single species multi-locus analysis, it should also be possible to integrate across different species sampled from the same locations within \pkg{BPEC}.  Such an approach would provide for quantitative measures for comparative phylogeography, and this will also be explored for future updates of \pkg{BPEC}.

Outgroup sequences can potentially directly inform about the probability of a haplotype being the Most Recent Common Ancestor (MRCA) of a set of sequences, and future versions of \pkg{BPEC} will explore the possibility of incorporating outgroup sequences to for this purpose. Sequences immediately derived from an inferred MRCA are also expected to provide some information regarding ancestral areas, and integrating information across the MRCA and sequences immediately derived from it will also be explored.
					
Finally, we are investigating whether we can extend the applicability of \pkg{BPEC} to the analysis of geographic population structure derived from vicariant processes - i.e.,~where populations become isolated and thus initially share genetic variation, but diverge through time through lineage sorting effects and the accumulation of new population specific mutations. \pkg{BPEC} should be applicable to the examination of genealogy among such closely related populations under the evolutionary model of population splitting. In the absence of opposing gene flow among populations, all populations will eventually become diagnosable as descending from a single haplotype unique to that population (lineage sorting). This diagnostic is equivalent to the pattern derived from a colonisation event, and as such it must be borne in mind that clusters defined by \pkg{BPEC} may indeed have a vicariant origin. Incorporating a vicariance model into \pkg{BPEC} may prove challenging, but it would (i) facilitate the detection of more subtle geographic structuring than that derived from the dispersal model, and (ii) provide a more realistic model of phylogeographic structure.

\section*{Acknowledgements}
We would like to thank the anonymous reviewers and associate editor for their thoughtful comments which have greatly improved the manuscript. We would also like to thank all the users who have provided feedback on \pkg{BPEC} over the years. AH thanks D.~Tarkhnishvili for his expert comments on the brown frogs. N.~Arrigo, T.~Nepusz, L.~Revell and G.~Valiente quickly answered questions on \proglang{R} functions they published, many thanks to them all. 

\bibliography{JSS2640.bib}

\clearpage
\section{Appendix}
\subsection{Temporal orderings}
\label{appendix:temporal}
\begin{figure}[h!]
\centering
\includegraphics[width=10cm]{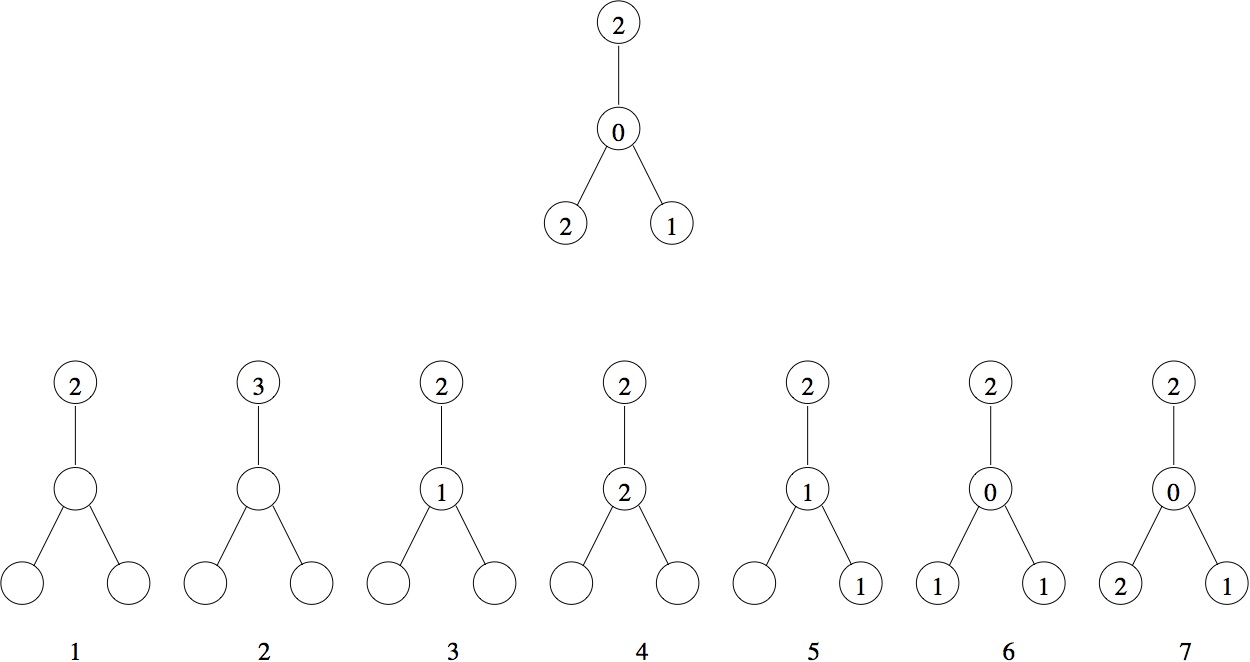}
\caption{\label{hproposalfigure}Top panel: In this tree the oldest haplotype of the sample  (the top haplotype) is observed twice in the sample, whereas the intermediate haplotypes in not observed at all.
Bottom panel: a possible time evolution of how the haplotypes arose. Nodes without a number correspond to haplotypes which haven't appeared yet. At first one sequence is present, the ancestral sequence, which replicated into two (the first event is always a replication, otherwise that haplotype would disappear). Then one of those two identical sequences may replicate again to give us a total of three (or could have mutated to give a new haplotype). One of those three then mutate to give us the intermediate haplotype, which in turn  here replicates and then mutates (and goes extinct) to give us the right-hand leaf. Finally, the intermediate haplotype mutates again to give us the left-hand leaf, which then also replicates to give another copy of itself. }
\end{figure}

Suppose the haplotype tree is given by the top tree of Figure~\ref{hproposalfigure} \citep{manolopoulou2012phylogeographic}. For ease of exposition, the numbers on the nodes here represent the sample sizes of each haplotype rather than the label of each haplotype. 

Simulating a temporal ordering implies that, starting with the ancestral sequence, we specify a series of replication and mutation events which occurred by mimicking evolution, eventually resulting in the observed haplotype tree. A possible series of events is shown in Figure~\ref{hproposalfigure} through 7 timepoints, where node numbers indicate number of copies of each haplotype. 

Notice that, if the root node had replicated further, we would have had three copies of the root haplotype. Although in theory this could have happened, with one of the copies eventually becoming extinct, we do not take into account any such scenarios, instead we only account for the observed sequences. Additionally, it would not have been possible for the intermediate haplotype to mutate after Step 3 above, since then it would disappear from the ancestral sequences, and another mutation would not have been possible.

\subsection{Computational issues}
\label{appendix:issues}
\subsubsection{Haplotype tree likelihood }
When calculating the Metropolis-Hastings ratio for a proposal from root $r$ to $r'$, one needs to calculate 
\begin{eqnarray}
\frac{p(r'\mid T,\mathcal{S})}{p(r\mid T,\mathcal{S})}=\frac{ \frac{\left|\mathcal{O}^\mathcal{S}_{r',T}\right|}{\sum_{r}\left|\mathcal{O}^\mathcal{S}_{r,T}\right|}}{ \frac{\left|\mathcal{O}^\mathcal{S}_{r,T}\right|}{\sum_{r}\left|\mathcal{O}^\mathcal{S}_{r,T}\right|}}=\frac{ \left|\mathcal{O}^\mathcal{S}_{r',T}\right|}{\left|\mathcal{O}^\mathcal{S}_{r,T}\right|}.\nonumber
\end{eqnarray}
However, computing the size of the two sets of temporal orderings is a computational bottleneck. To overcome this issue, an unbiased estimator of the likelihood is used instead. Conditionally on a root $r$ and tree $T$, a particular ordering $O^*$ is generated by moving from the root to the tips and randomly choosing among the available replicate/mutate moves at each step, according to some distribution $q(O^*)$, such that any possible ordering of $\mathcal{O}^\mathcal{S}_{r,T}$ can arise. Then calculate 
\begin{eqnarray}
\widehat{\left|\mathcal{O}^\mathcal{S}_{r',T}\right|}&=& \frac{1}{q(O^*)}\nonumber,
\end{eqnarray}
such that 
\begin{eqnarray}
\mathbb{E}\left(\widehat{\left|\mathcal{O}^\mathcal{S}_{r',T}\right|}\right)&=& \sum_{O^*\in\mathcal{O}^\mathcal{S}_{r,T} } \frac{1}{q(O^*)}\times q(O^*)= \left|\mathcal{O}^\mathcal{S}_{r',T}\right|\nonumber,
\end{eqnarray}
so it provides an unbiased estimator of the likelihood. Note here that the latent variable $O$ is not accepted/rejected together with the root, so the Markov chain Monte Carlo does not maintain detailed balance \citep[see][]{manolopoulou2012phylogeographic,beaumont}. This is because variance of $q(O^*)$ can be huge and  detrimental to the MCMC, causing it to get `stuck';  future improvements of $q(O^*)$ could allow $O^*$ to be accepted/rejected together with the root $r$. Multiple realizations of $O^*$ could also be used instead, but \pkg{BPEC} only considers 1. 

\subsubsection{MCMC exploration and convergence}
The \pkg{BPEC} model faces two additional key computational bottlenecks. The first comes from learning the posterior probability of the root haplotype. Since it relies upon an estimator of the likelihood, a large number of iterations are required in order to allow for reasonable convergence. However, the total number of haplotypes (and as such the number of possible roots) is generally low (usually up to a few hundreds), so with enough iterations the sampler can explore the whole root parameter space sufficiently. 

On the other hand, the clustering parameter space is challenging to adequately explore. Instead, sophisticated local proposals are required. \cite{manolopoulou2011bayesian} implement a clustering proposal which cumulatively adds observation branches (as shown in Figure~\ref{haplotreesubdivided}) to clusters by starting with empty clusters with mean and variance equal to their corresponding prior means. As each observation branch is added to one of the clusters (in random order), the means and variances of that cluster are updated according to the corresponding posterior means. This allows the sampler to propose clusters for each branch according to the cluster in which it fits best, while randomising the order of the allocation meant that no branches were given higher weight than others. 
In \pkg{BPEC} we tweak the proposal distribution of \cite{manolopoulou2011bayesian} by introducing an auxiliary variable $w_c$, representing the weight of the previous clustering in the MCMC sampler. Rather than simply allocating each observation branch to one of the existing clusters simply by assessing the fit of each branch to each of the clusters, we assign it to the same cluster as the previous iteration (where possible) with probability $w_c$. This favours clusterings similar to the previous iteration, thereby ensuring that local moves are proposed more frequently. Since $w_c$ is an auxiliary variable, it is accepted/rejected together with the proposed parameters, so the sampler automatically chooses a value of $w_c$ that is reasonable. 
\subsubsection{Label-switching}
In order to draw cluster-specific inferences, cluster labels need to be assigned for every posterior sample available. This is known as the label-switching problem \citep{stephensmixture,labelling,papastamoulis2010artificial}
and it is especially challenging in the case of a variable number of clusters. Here we take a pivoting approach to assign cluster labels on-line (i.e.,~without the need of post-processing). The algorithm works as follows: 
\begin{enumerate}
\item During burn-in of the first chain, record the cluster labels of the posterior sample with the highest value of the posterior density, denoted by $\textbf{c}^*$. 
\item Once this clustering is fixed, subsequent labels of the posterior sample of the set $\left(\boldsymbol{\mu},\boldsymbol{\Sigma}\right)$
are chosen such that $$p\left(\mathbf{Y}\mid \boldsymbol{\mu}_{c^*},\boldsymbol{\Sigma}_{c^*},c^*\right)$$ is maximised. 
In other words, labels of the set of means and covariances are chosen such that the likelihood relative to (approximate) maximum a posteriori clustering $c^*$ is maximised. 
\end{enumerate}
\subsubsection{Hashing}\label{hashlabelling}
In contrast to coalescent trees, which are binary and can be represented simply by the pairs of subsequent coalescence events, haplotype trees do not have shorthand representations. Instead, a standard way to represent a haplotype tree is through its corresponding graph adjacency matrix. However, keeping track of posterior samples of trees requires storing entire matrices at each iteration of the sampler, which creates a memory bottleneck. 

In our case, we can take advantage of the fact that not all adjacency matrices are possible; most edges are either certainly present or absent as determined by $\Omega$. Uncertainty only arises through edges that are part of a loop in the network, so each tree is characterised by the set of deleted edges. Trees are then reduced to vectors of length $n_{loop}$ with integer entries. Standard hashing techniques can thus be used to store the number of times each tree (i.e.,~each integer vector) appears in the MCMC posterior samples. 

Hashing algorithms allow us to represent integer vectors by an single integer. In our case, we can store the index of the edge deleted from each loop at each iteration of the MCMC, keeping track of them via the `hashing index' of the entire vector. 
Hash functions create a short (as short as possible) address book where each of these numbers is stored in a specific page, in such a way that it can easily be retrieved \citep[see][]{hash1}.
\end{document}